# High Sensitivity Biosensor using Injection Locked Spin Torque Nano-Oscillators

Tathagata Srimani[1], Bibhas Manna[1], Anand Kumar Mukhopadhyay[1], Kaushik Roy[2], Mrigank Sharad[1]
[1]Department of Electronics and ECE, IIT Kharagpur, West Bengal, India
[2]School of Electrical Engineering, Purdue University, West Lafayette, IN, USA

## ABSTRACT

With ever increasing research on magnetic nano systems it is shown to have great potential in the areas of magnetic storage, biosensing, magnetoresistive insulation etc. In the field of biosensing specifically Spin Valve sensors coupled with Magnetic Nanolabels is showing great promise due to noise immunity and energy efficiency [1]. In this paper we present the application of injection locked based Spin Torque Nano Oscillator (STNO) suitable for high resolution energy efficient labeled DNA Detection. The proposed STNO microarray consists of 20 such devices oscillating at different frequencies making it possible to multiplex all the signals using capacitive coupling. Frequency Division Multiplexing can be aided with Time division multiplexing to increase the device integration and decrease the readout time while maintaining the same efficiency in presence of constant input referred noise.

## 1.0 INTRODUCTION

The advancements on genomics and proteomics has ushered a new era in medical diagnostics specially in molecular medicine with future possibilities in personalised therapeutic techniques. This greatly relies on the efficient and error free detection and sequencing of essential biomolecules like DNA, RNA, proteins etc. Presently biomolecular detection is based on optical techniques .With the advancement of nanotechnology however single molecule detection using electronic devices has been made possible. Novel platforms based on GMR SV sensors with in situ CMOS architecture for signal transmission has already been proposed. Here the presence of biomolecules is transduced to a change in electrical resistance. In our proposed structure we are using TMR based sensors with the presence of a biomolecule being translated in the output voltage leading to easy noise free detection. The main advantage of magnetic detection can be attributed to low levels of magnetic residues in clinical samples giving rise to low noise and efficient detection.

In this scheme (Figure. 1) a DNA sample, is immobilised on the surface of the sensor through a complementary or anti sense sample. The DNA samples are biotinylated at the beginning and then introduction of this sample, which may contain hundreds to thousands of different DNA strands in different concentrations, the anti sense strand selectively binds only with the targeted DNA. The sample is incubated and the unbound DNA or other biomolecules are washed away. After a short interval, magnetic nanoparticles are added which attach to the sense strand through a high affinity biotin-streptavidin bond to complete the assay. Each magnetic nanoparticle is around 200 nm in diameter. This can be extended to detection of other essential biomolecules such as RNA, proteins etc where complementary DNA can be replaced by a c-RNA or capture antibody particular to the type of biomolecule.

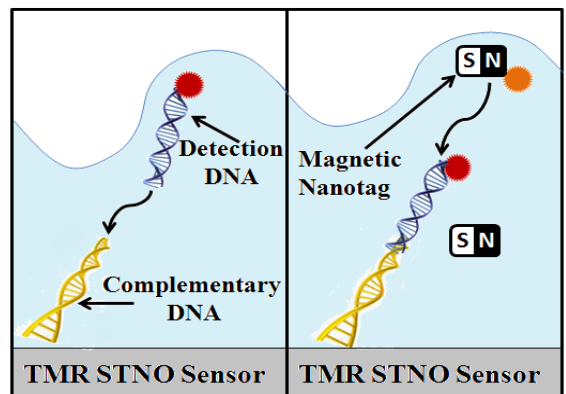

Figure 1: Schematic Model for STNO based DNA sensor. Detection DNA selectively binds with complementary DNA along with the Magnetic Nanotag leading to change in amplitude of the signal.

## 2.0 LITERATURE REVIEW

Biosensors act as a bridge between biology and instrumentation by converting a biological or chemical response into an electrical signal. Biosensors take on many different forms and sensing modalities, but can be broadly classified into two categories: labeled and label-free techniques. Labeled techniques tag a recognition antibody which binds to the biomolecule of interest with an externally observable label such as a fluorophore[3], quantum dot[4], electrochemical tag [5], or magnetic tag [6]–[10]. On the other hand, label-free techniques detect an intrinsic propertyof the biomolecule, such as the mass [11], charge [12], thermal reactivity [13], size [14], or its optical interaction with a surface [15]. This work concentrates specifically on magnetic labels that offer several key advantages over conventional optical techniques and other competing sensing methods. First, the samples (blood, urine, serum,etc.) naturally lack any detectable magnetic content, providing a sensing platform with a very low background signal. This is a significant and fundamental problem with many optical techniques where one encounters label-bleaching [16] and auto fluorescence [17]. Second, the sensors can be arrayed and multiplexed to perform complex protein or nucleic acid analysis in a single assay without resorting to bulky optical scanning. Additionally, the magnetic tags can be manipulated with a magnetic field to potentially speed up the reaction [18] or remove unbound tags [19]. Finally, the sensors are compatible with CMOS IC technology, allowing them to be manufactured with integrated electronic read out, produced in mass quantities (potentially at low cost), and deployed in a one-time use, disposable format for point-of-care testing. Currently GMR based Spin Valve sensors have been reported in literature [1,2] which exploits a quantum mechanical effect where change in magnetic flux leads to change in electrical resistance due to spin dependent scattering. But the major drawback of such sensing scheme is the low output voltage which demands a complex circuitry for amplification and signal processing. In the scheme proposed in this report, TMR based spin torque nano oscillators are used for detecting magnetically tagged biomolecules. A change in the bias magnetic field leads to change in the oscillation amplitude which is translated to change in electrical resistance leading to detection. This sensing scheme offers a significant advantage over the previously reported GMR SV sensors owing to its high sensitivity and high output voltage leading to a relatively simpler analog frontend for detection.

### 3.0 PROPOSED STO BASED BIOSENSOR

### 3.1 DEVICE STRUCTURE

The magnetic sensor being used is basically a two terminal Spin Torque Nano Oscillator (STNO) which has a thin insulating oxide sandwiched between two ferromagnetic layers (Tunneling Magneto Resistance-TMR) (Figure 2b). The ferromagnetic layers have two stable spin-polarization states, depending upon magnetic anisotropy. The magnetization of one of the layers is fixed, while that of the other (free-layer) can be influenced by a charge current passing through the device or by an applied magnetic field. The high-polarity fixed magnetic-layer spin-polarizes the electrons constituting the charge-current, which in turn exert spin transfer torque (STT) on the free layer. STT results in precession of the effective spin moment of the free layer away from the original magnetization axis, while an inherent inertial damping torque tries to restore it along the original state.[ref] A static magnetic field can be used to obtain sustained spin-precession of the free layer at an angle $\varphi$, at which the STT and the damping torque balance out each other. The dynamics of the Free Layer is governed by the Landau-Lifshitz-Gilbert Equation

$$\frac{dm}{dt} = -|\gamma| m \times H_{eff} + \alpha \left( m \times \frac{dm}{dt} \right) + |\gamma| \left| \frac{h}{\mu_0 e} \right| \frac{J}{t_f M_s} P(m \times m_p \times m) \qquad (1)$$

which is composed of three terms-precession term from the magnetic field, Spin Transfer Torque term from the current and the inherent damping term. (Figure 2b) In this equation, $\gamma$ is the gyromagnetic ratio, $\alpha$ is the damping constant, $h$ is the Plank's constant, $t_f$ is the free layer thickness, $M_z$ is the saturation magnetization of the magnet, $P$ is the polarization constant and $m_p$ is the spin polarization of the fixed layer. The resistance of the MTJ can be expressed as a function of relative angle θ between the spin-polarization of the two ferromagnetic layers as

$$R(\theta) = \left( \frac{R_P + R_{AP}}{2} \right) + \left( \frac{R_{AP} - R_P}{2} \right) \cos\theta \qquad (2)$$

where RP and RAP denote the resistance when the two layers are parallel and antiparallel. The absolute resistance of a GMR device is much smaller than that of a TMR device. A GMR based STNO, being fully metallic, can be operated with very low voltage (<10mV). However, the sensed signal amplitude is very low, which requires complex sensing circuitry to amplify the signal, leading to high power consumption. On the other hand, though the TMR based STNO can provide large amplitude output signals, due to the high resistance tunnel junction and thus can be suitable for signal analysis and also it can be seen later that the sensitivity of the TMR based sensors are higher than GMR sensors. The device dimensions and constants provided in (Table 2) has been used in rest of the paper, unless stated otherwise.

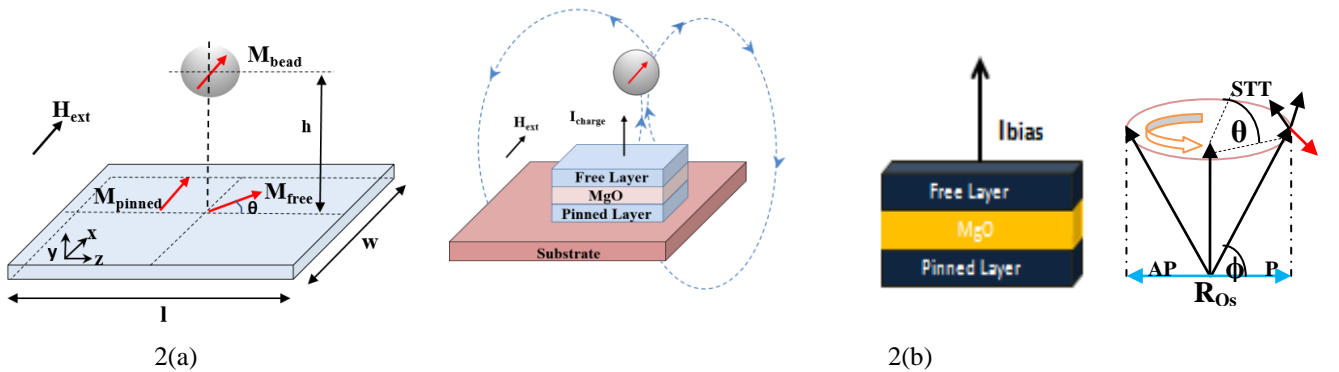

2(a)  2(b)

Figure 2:(a)Dimensions and pictorial representation of the nanobead with STO,(b) Schematic model of STNO and precession of the free layer components with different torque component

Table 2 Magnet and LLGS equation's parameters

| Parameter Value |
|---|
| Effective device area ($l \times w$): 30 x30 nm$^2$ |
| Free layer thickness ($t_F$): 1.5nm |
| Pinned layer thickness ($t_P$): 2nm |
| Spacer layer thickness ($t_{SP}$): 2nm |
| Gilbert damping coefficient ($\alpha$): 0.01 |
| Gyromagnetic ratio ($\gamma$): 2.21x10$^5$ rad.s$^{-1}$.T$^{-1}$ |
| Saturation Magnetization ($M_{SF}$)- 8x10$^5$ A/m |
| Saturation Magnetization ($M_{SP}$)- 15x10$^5$ A/m |
| Energy Barrier ($E_b$)- 40kT |
| Magnetic field : Static- (Hz): 5kOe |
| Time varying- (Hx): 10kOe* cos (0.7π/180) |
| Radius of nanobead: 200 nm |
| Distance of nanobead from sensor: 400nm |

## 3.2 ANALYTICAL MODEL

The magnetic nanoparticle which is introduced at the top of the sensor surface gets magnetised due to the external field ($\vec{H}_{ext}$) and the dipolar fields ($\vec{H}_{P/F}$) from the pinned and free layers of the Spin Torque Oscillator-

$$\vec{H}_{TNP} = \vec{H}_{ext} + \vec{H}_{P/F} \tag{3}$$

$$\vec{H}_{P/F} = \oint_{circumference} \frac{\vec{\sigma}_{surf} \vec{r}}{r^3} dl \tag{4}$$

where $\vec{\sigma}_{surf}$ is surface magnetic charge density, $\vec{r}$ is the centre to centre distance from the pinned and free layers to the nano bead. The polarization of the bead (Fe$_3$O$_4$) with saturation magnetisation $m_s = 480 emu/cc$ at a given temperature T can be expressed using the Langevin function as $m = m_s \left[ \coth\left(\frac{m_s H_{TNP}}{kT}\right) - \frac{kT}{m_s H_{TNP}} \right]$, to determine the dipole field of the bead averaged over the MTJ sensor which can be calculated as

$$\vec{H} = \frac{1}{lwt} \int_{-l/2}^{l/2} \int_{-w/2}^{w/2} \int_{-t/2}^{t/2} \left[ \frac{3(\vec{m}.\vec{r})}{r^5} - \frac{\vec{m}}{r^3} \right] dx dy dz \tag{5}$$

This magnetic field if high enough can alter the amplitude of oscillations of a STNO substantially, leading to detection of the nanobead. Magnetic mono domain simulation solving the LLG equation can estimate the frequency and the amplitude of this oscillations predicting the detection of the nanoparticle.

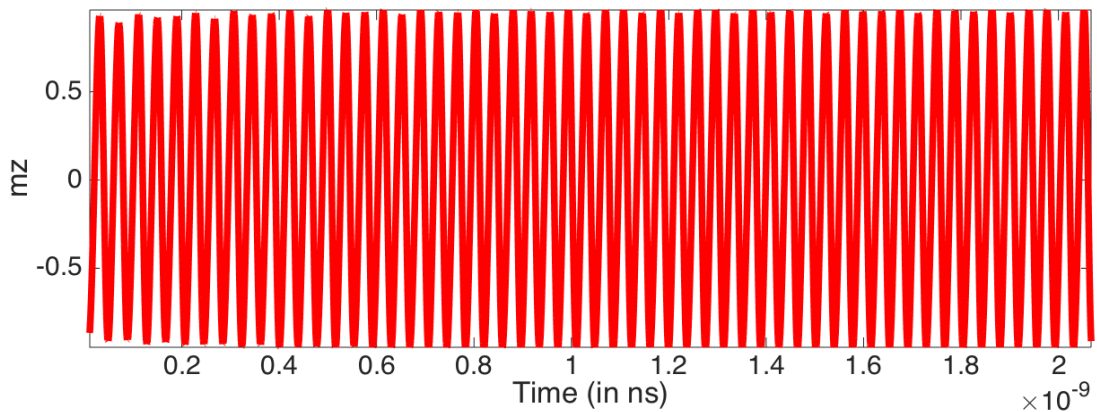

Figure 3: Oscillation of z component of free layer magnetization vs time

(Figure 3.a) shows the transient magnetization oscillation obtained by micro-magnetic simulation. The frequency of oscillation can be controlled by the magnitude of injected bias current (Figure 4a,b).It is evident from (Figure 6) that larger area STO requires more DC bias current to oscillate at the same frequency thus consuming more power. Again a static magnetic field (Figure 4a) applied induces a smaller range of frequency upon variation of the current unlike a combined static and time varying magnetic field (Figure 4b) which may lead to larger range of operable frequencies under the same range of dc bias current.

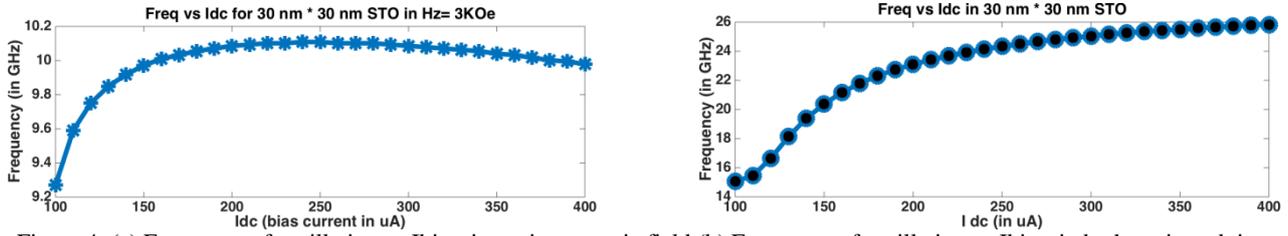

Figure 4: (a) Frequency of oscillation vs Ibias in static magnetic field (b) Frequency of oscillation vs Ibias in both static and time varying field

Now the oscillation of the magnetisation induces a change in (theta) leading to oscillation in the resistance as well.The amplitude of these oscillations however strongly depends on the applied magnetic field.Thus any change in the magnitude of the magnetic field may lead to significant change in (theta) leading to change in maximum and minimum resistance. Thus the presence of a magnetic nanoparticle may cause significant change in static magnetic field which is translated to change in voltage levels in the output. To establish error free operation of multiple sensors we need to quantify the device parameters and make the system immune to random noise due to thermal fluctuations and noise from detection circuitry. This calls for the need to lock each of the individual sensors to a particular frequency of operation which can be achieved by application of an RF signal of small magnitude. In order to establish locking, a common ac current signal can be injected into a larger number of oscillators. If the external ac frequency is close to that of the bias frequency of the STOs (determined by the dc bias), they phase lock to the injected ac current signal. Higher rfcurrent establishes stronger locking and increase in amplitude of oscillations. (Figure 5)

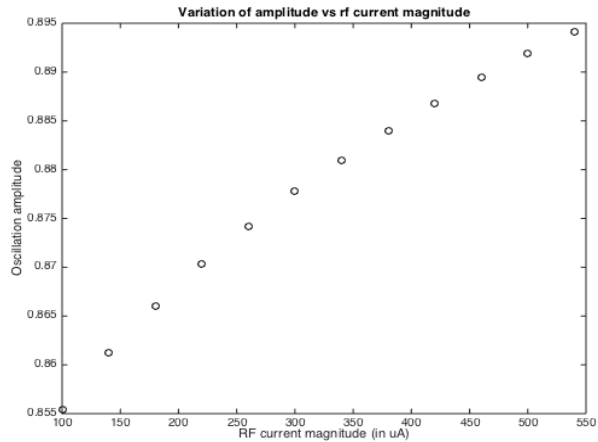

Figure 5: Amplitude of oscillation vs RF current Amplitude

Also the individual sensors need to be separated by a frequency margin so as to decrease the effect of noise and this margin can be effectively determined by calculation of the rms value of the deviation angle theta which in turn can be obtained using equipartition theorem [ref] (provide equation)

$$\sin(\theta_{rms}) = \left(\frac{k_B T}{2E_b}\right)^{\frac{1}{2}} \qquad (6)$$

The rms value of the normalized component of magnetization about its easy axis is calculated using $(m_Z)_{rms} = \cos(\theta_{rms})$ where $k_B$, $T$ and $E_b$ are the Boltzmann constant, temperature and effective anisotropy energy barrier. The volume of the sensor is kept fixed while computing the $(m_Z)_{rms}$ for higher $E_b$ indicating the use of magnetic materials with larger uniaxial magnetocrystalline anisotropy (~3x10$^4$ J/m$^3$) which is achievable using monodisperse FePt nanoparticles with controlled size and composition.This calibrated noise when included in the LLG equation gives rise to statistical variation of the oscillation frequency which can be captured using Monte Carlo simulation on the oscillator (Figure 7).

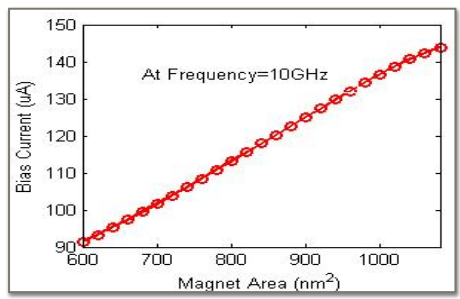

Figure 6: Bias current vs Area for same oscillation frequency (10 GHz)

The frequency margin is obtained for a dc bias current of 200uA and the sensor is injection locked at the same frequency imposed by the bias. The simulation results in the statistical estimate of the frequency margin as 0.1GHz.If current resolution of 10uA is considered the above estimate can lead to integration of 20 such devices keeping the issue of power in mind.

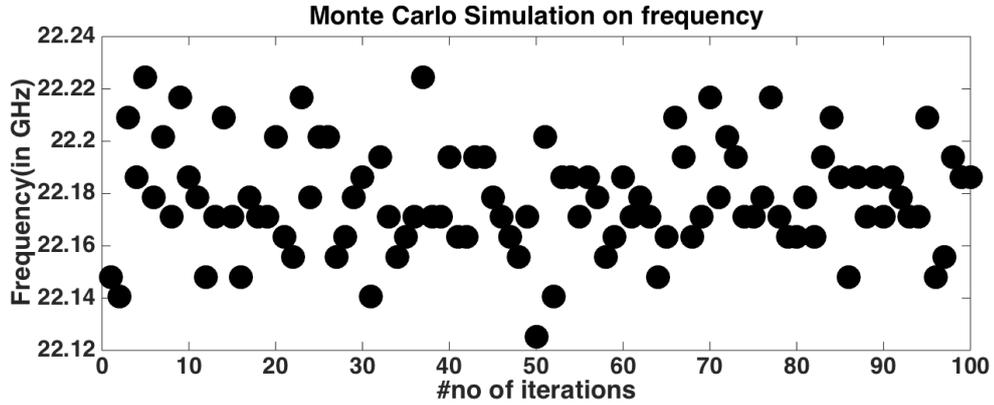

Figure 7: Monte Carlo Simulation of frequency in presence of noise

External magnetic field (~5kOe) is applied along x direction and the frequency of oscillation is locked at different frequencies by injecting rf current as shown in (figure). It is seen that the frequency of oscillation decreases at higher energy barrier which is due to the increase in uniaxial in-plane anisotropy field. Next a magnetized MNP with 200 nm diameter is introduced at a height of 400 nm from the sensor surface to observe the change in amplitude of oscillation as shown in (Figure 11c) where the dc bias current is taken to be 200 uA and corresponding rf locking current optimised to maximise the sensitivity of the sensor.(Figure 8).Higher rf current imposes better injection locking leading to magnetic field having less pronounced effect which in turn decreases the sensitivity.

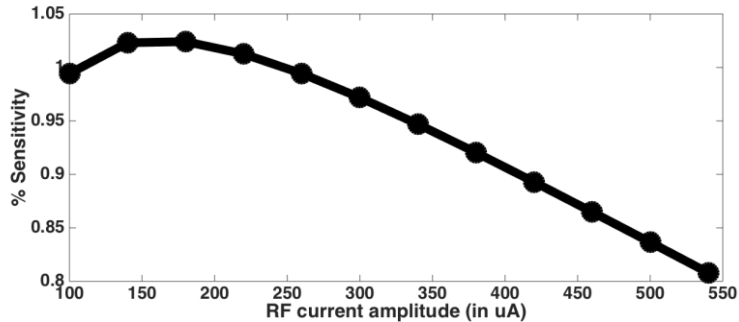

Figure 8: Sensitivity vs RF current magnitude (Sensitivity defined as $\frac{\Delta A}{A0}$ where *A0* is the amplitude without the bead and *ΔA* the change in amplitude with introduction of the bead.*This definition will be followed throughout the paper unless stated.)

An analyses on external field shows (Figure 9a,b) that the change of oscillation amplitude rather than the oscillation frequency is a strong function of the magnetic field thus making it a suitable parameter for sensing.

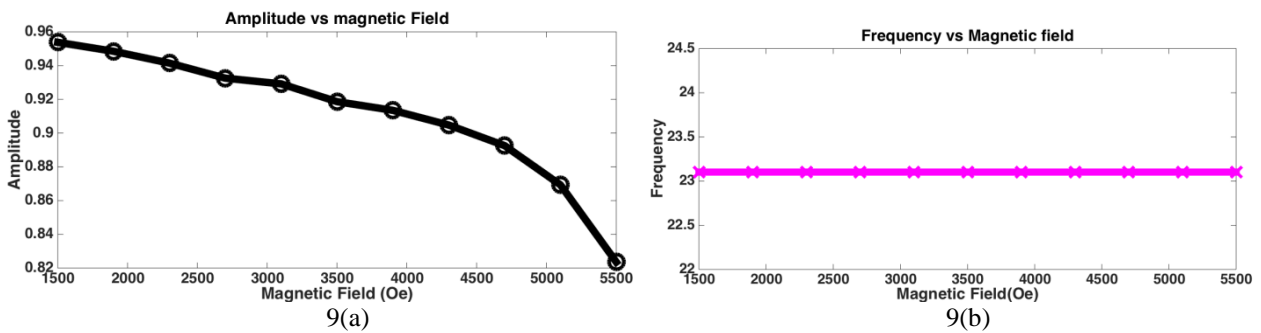

9(a)          9(b)

Figure 9: (a) Amplitude vs Magnetic field (b) Frequency vs Magnetic field

**4.0 RESULTS & DISCUSSIONS:**

Figure10 shows the variation of sensitivity with variation in device parameters such as Ea, magnet thickness, alpha and MNP parameters - radius and distance from the surface. It can be observed that bead of higher radius placed closer to the surface generates maximum sensitivity which is due to the increase in the stray field opposing the static magnetic field. Similarly higher Ea increases stability of oscillations thus decreasing the sensitivity unlike higher damping factor alpha which leads to increased instability and thus higher sensitivity.

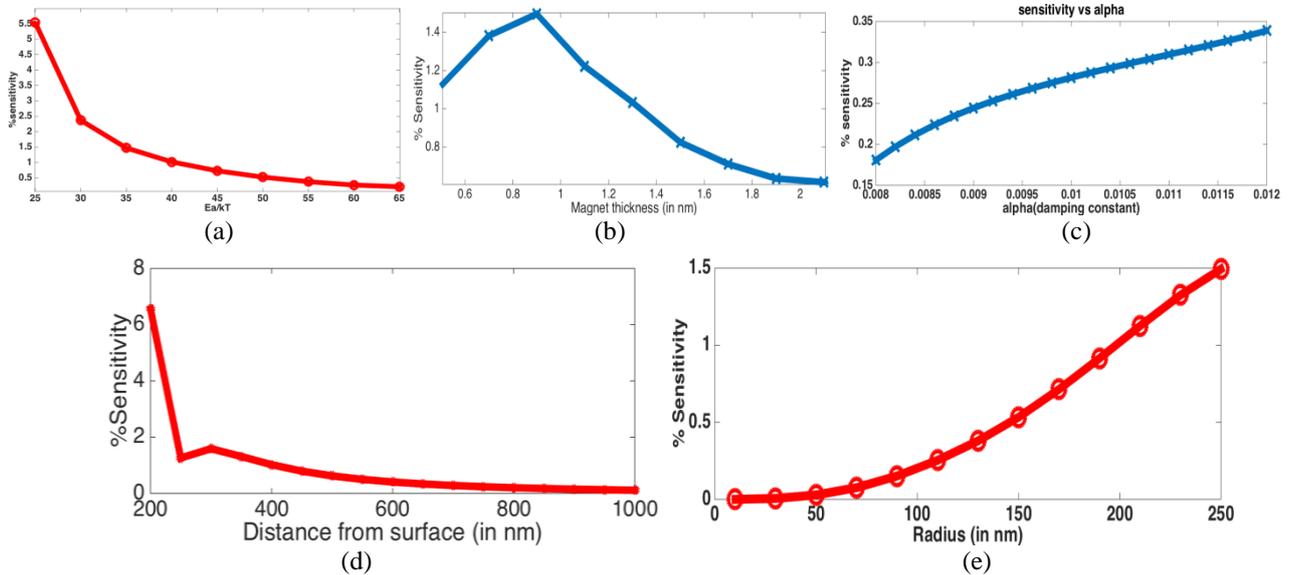

Figure 10: (a) Sensitivity vs Ea (b) Sensitivity vs Magnet thickness (c) Sensitivity vs alpha (d) Sensitivity vs distance of bead from surface (e) Sensitivity vs bead size

Again, change in Ms (saturation magnetisation) and alpha (damping constant) can cause a significant change in oscillation frequency of each oscillator (Figure 11c, d) leading to poor locking and decrease in amplitude (Figure 11a, b) and decrease in sensitivity. In order to take this effect in account we propose a conceptual calibration setup (Figure 11e). This can be used to tune the dc bias current which in turn brings the oscillator back to the desired injection locked frequency.

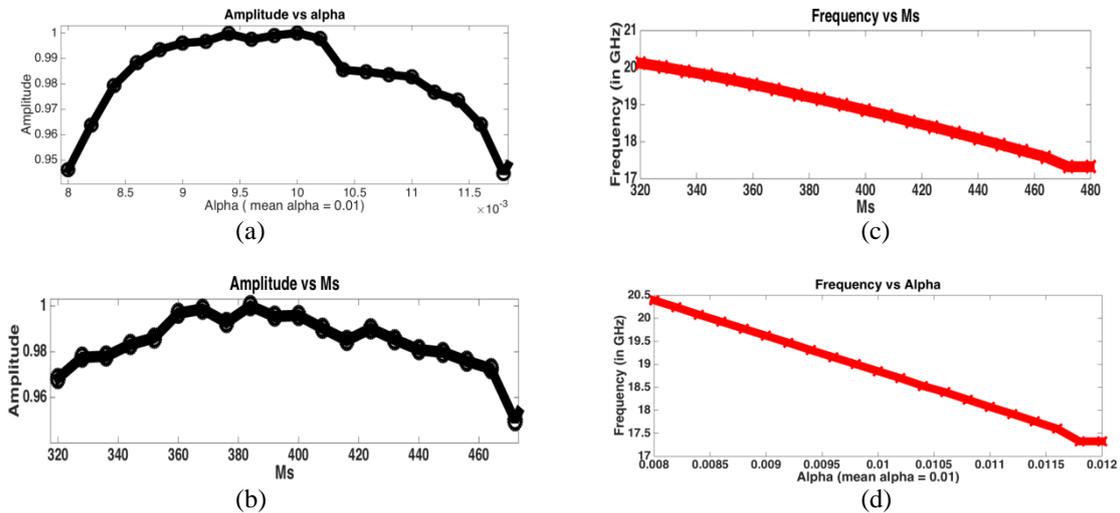

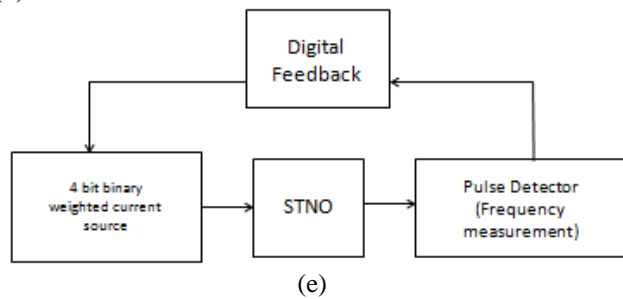

Figure 11: (a) Amplitude vs Ms (b)Amplitude vs alpha (c) Frequency vs Ms (d)Frequency vs alpha (e) Conceptual calibration setup for tuning the dc bias. *Device parameters : - Ea = 40kT Area = 900 nm$^2$ , calibrated noise taken in account

Optimising all these parameters for maximum sensitivity 20 such frequency locked devices are dynamically modeled (Figure12a, b) based on the noise considerations. With a separation of at least 0.1GHz and dc bias resolution of 10 uA . The average sensitivity (figure) is around 1% (Figure 12c) which is a significant improvement on GMR based sensors and also here it may be possible for detection of single biomolecule.

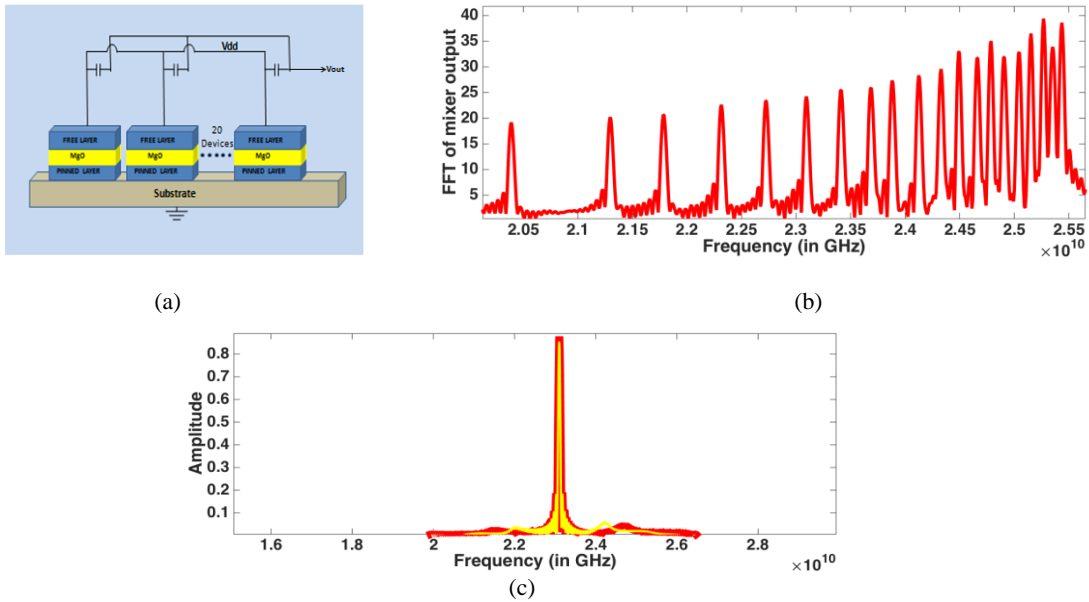

(a)   (b)

(c)

Figure 12: (a) Proposed device architecture (b) FFT of mixed signal output (c) FFT of output signal from a STO with(yellow) and without nanobead (red) change in amplitude is ~1% signifying detection

Frequency division multiplexing can be assisted by time division multiplexing to increase number of sensors for large area detection where the sensor array can feed the output signal to an ADC connected to a digital signal processor for low cost on chip data analysis.

**5.0 CONCLUSION**

We proposed and simulated a sensor system model based on STNO using micro magnetic mono domain simulation and we observed and analysed the effect of parameter variation on the sensitivity 2T- STO based sensor. We proposed a sensor system based on the integration of multiple sensors for frequency division multiplexing with optimisation of various parameters for maximum sensitivity. The proposed device shows a significant reduction of area where the individual sensors have dimensions of 30 nm X 30 nm compared to 120 um X 120 um in GMR SV sensor array [1] and also the number of devices integrated per chip can be enhanced due to more number of frequency channels compared to 4 reported previously. Also each of the STNOs consumes power in the order of microwatts thus the net power consumed per chip will be around few miliwatts for 1000 sensors in a single chip. Thus this proposed architecture can be highly robust, energy efficient and easily integrable with a simple CMOS circuitry for analysis and detection and can be extended to a general scheme for detecting target biomarkers and essential biomolecules.

**REFERENCES:**

**[1]** Drew Hall et. al, IEEE Journal of Solid State Circuits, vol.48, no.5, May 2013

**[2]** Drew Hall et. al., Circuits and Systems (ISCAS), Proceedings of 2010 IEEE International Symposium on, ISCAS 2010, pp -1779-1782

**[3]** R.Curry, H.Heitzman, D.Riege, R.Sweet, and M.Simonsen, Clin.Chem., vol.25, no.9, pp.1591-1595, Sep.1979.

**[4]** M.Han, X.Gao, J.Z.Su, and S.Nie, Nature Biotech., vol.19, no.7, pp.631–635, Jul.2001.

**[5]** P.M.Levine, P.Gong, R.Levicky, and K.L.Shepard, IEEE J. Solid-State Circuits, vol.43, no.8, pp.1859–1871, Aug. 2008.

**[6]** R.S.Gaster, D.A.Hall, C.H.Nielsen, S.J.Osterfeld, H.Yu, K.E.Mach, R.J.Wilson, B.Murmann, J.C.Liao, S.S.Gambhir, and S.X.Wang, NatureMed.,vol.15,no.11,pp.1327–1332,Nov.2009.

**[7]** H.Wang, Y.Chen, A.Hassibi, A.Scherer, and A.Hajimiri, in IEEE Int. Solid-State Circuits Conf. Dig. Tech. Papers, ISSCC 2009, 2009, pp. 438–439, 439a.

[8] H.Lee, E.Sun, D.Ham, and R.Weissleder, NatureMed., vol.14,no.8, pp.869–874, Jul.2008.

[9] O.Florescu, M.Mattmann, and B.Boser, J.Appl. Phys., vol. 103, no.4, p.046101, 2008.

[10] P.Liu, K.Skucha, Y.Duan, M.Megens, J.Kim, I.Izyumin, S.Gambini, and B.Boser, in Symp. VLSI Circuits (VLSIC) Dig., 2011, pp.176–177.

[11] J.Fritz, M.K.Baller, H.P.Lang, H.Rothuizen, P.Vettiger, E.Meyer, H.J.Güntherodt, C.Gerber, and J.K.Gimzewski, Science, vol.288, no.5464, pp.316–318, Apr.2000.

[12] E.Stern, J.F.Klemic, D.A.Routenberg, P.N.Wyrembak, D.B.Turner Evans, A.D.Hamilton, D.A.LaVan, T.M.Fahmy, and M.A.Reed, Nature,vol.445,no.7127,pp.519–522,Feb.2007.1300

[13] C.Hagleitner, A.Hierlemann, D.Lange, A.Kummer, N.Kerness, O.Brand, and H.Baltes, Nature, vol.414, no.6861, pp. 293–296, Nov.2001.

[14] J.Clarke, H.-C.Wu, L.Jayasinghe, A.Patel, S.Reid ,and H.Bayley, Nature Nano, vol.4, no.4, pp.265–270, Apr. 2009.

[15] C.Boozer, G.Kim, S.Cong, H.Guan, and T.Londergan, Current Opinion Biotechnol., vol.17, no.4, pp.400–405, Aug.2006.

[16] H.Gilohand J.Sedat, Science, vol.217, no.4566, pp.1252–1255, Sep.1982.

[17] J.E.Aubin, J.Histochem. Cytochem.,vol.27, no.1, pp.36–43, Jan.1979.

[18] D.M.Bruls, T.H.Evers, J.A.H.Kahlman, P.J.W.vanLankvelt, M.Ovsyanko, E.G.M.Pelssers, J.J.H.B.Schleipen, F.K.deTheije, C.A.Verschuren, T.vanderWijk, J.B.A.vanZon, W.U.Dittmer, A.H.J.Immink, J.H.Nieuwenhuis, and M.W.J.Prins, LabChip, vol.9,no.24, pp.3504–3510, Oct.2009.

[19] T.Aytur, J.Foley, M.Anwar, B.Boser, E.Harris, and P.R.Beatty, J. Immunolog. Meth., vol.314, no. 1–2, pp.21–29, Jul. 2006.